\documentclass[preprint,showpacs,superscriptaddress]{revtex4}

\usepackage{graphicx}

\begin{document}

\title{COMMUNITY STRUCTURE IN JAZZ}

\author{PABLO M. GLEISER}

\affiliation{Departament de F\'{\i}sica Fonamental, Universitat de
Barcelona, Diagonal 647,\\ 08028 Barcelona, Spain.
gleiser@ffn.ub.es.}

\author{LEON DANON}

\affiliation{Departament de F\'{\i}sica Fonamental, Universitat de
Barcelona, Diagonal 647,\\ 08028 Barcelona, Spain.  ldanon@ffn.ub.es.}

\begin{abstract}
  Using a database of jazz recordings we study the collaboration network
  of jazz musicians. We define the network at two different
  levels. First we study the collaboration network between individuals,
  where two musicians are connected if they have played in the same
  band. Then we consider the collaboration between bands, where two
  bands are connected if they have a musician in common. The community
  structure analysis reveals that these constructions capture essential
  ingredients of the social interactions between jazz musicians. We
  observe correlations between recording locations, racial segregation
  and the community structure. A quantitative analysis of the community
  size distribution reveals a surprising similarity with an e-mail based
  social network recently studied.
\end{abstract}


\maketitle

\section{Introduction}    

In the last years the physics community has devoted a strong effort to
the study of social networks \cite{Newman1}. The availability of large
databases containing information on the collaborations between movie
actors, scientists, etc.  has allowed for many statistical properties
of these networks to be characterized. \cite{Albert}.  They revealed
that some characteristics appear to be general for these kind of
networks.  In particular they show the so called small world property,
that is the average distance between vertices is small, while the
clustering of vertices remains high.It has also been observed that the
degree distribution $P(k)$ is skewed. In the particular case where
$P(k)$ presents a power law tail the network is known as scale free.\cite{Barabasi}.

An interesting point which has recently been brought to attention is
the community structure of networks. Communities appear in networks
where vertices join together in tight groups that have few connections
between them.  By eliminating these connections it is possible to
isolate the communities. In fact this is the main idea behind the
algorithm that Girvan and Newman have recently proposed \cite{Girvan}.

In this work we study the topology and the community structure of the
 collaboration network of jazz musicians. In order to do so we
 construct the network at two different levels.  First we build the
 network from a 'microscopic' point of view.  In this case each vertex
 corresponds to a musician, and two musicians are connected if the
 have recorded in the same band. Then we build the network from a
 'coarse-grained' point of view. In this case each vertex corresponds
 to a band, and a link between two bands is established if they have
 at least one musician in common. Clearly, this is the most simple way
 in which one can establish a connection between bands, and the
 definition can be extended to incorporate directed and/or weighted
 links. However, we show that even by using this simple definition we
 are able to recover essential ingredients of the social interactions
 between the musicians.

We characterize the topology of these networks by studying the degree
 distributions, clustering and average nearest neighbors degree. A
 comparison of the quantitative properties of the musicians and bands
 networks reveals differences which are analyzed in detail. We also
 study the community structure of both networks.  This analysis
 reveals the presence of communities which have a strong correlation
 with the recording locations of the bands, and also shows the
 presence of racial segregation between the musicians. These
 characteristics are well documented in jazz history \cite{Grove}.

Finally, we characterize quantitatively the community size
distribution $P(s)$. The shape of this distribution presents a
surprising similarity with the results reported in a recent study of
the e-mail network in the University Rovira i Virgili in Tarragona,
Spain \cite{Guimera}.

\section{The Jazz Network}

\begin{figure}[!hb]
\vspace{0.4cm}
\centerline{\includegraphics[width=0.65\columnwidth]{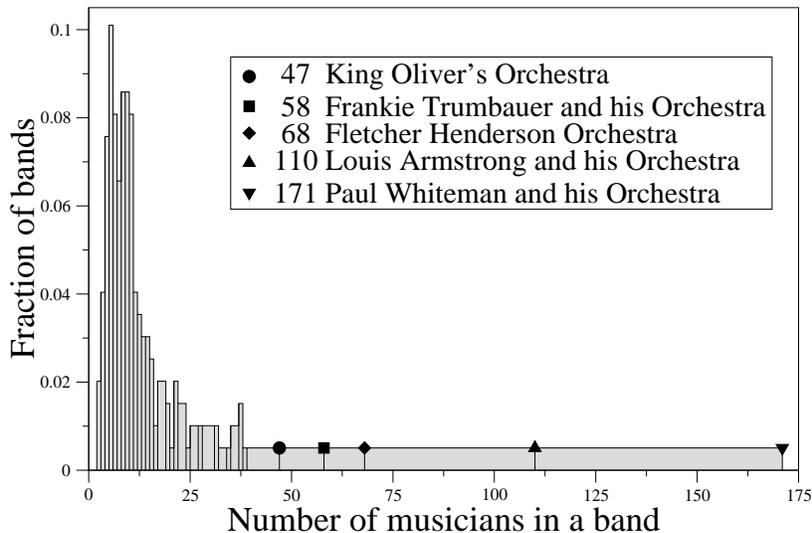}}
\vspace*{8pt}
\caption{\label{fig1} Distribution of musicians per band. The
horizontal axis corresponds to the number of musicians that have
played in a given band, while the vertical axis presents the fraction
of bands with the given number of musicians. The inset shows the names
of the five bands that include the largest number of musicians.}
\end{figure}

The data was obtained from {\it The Red Hot Jazz Archive} digital
database \cite{redhotjazz}.  In our analysis we include $198$ bands
that performed between $1912$ and $1940$, with most of the bands in
the $1920$'s.  The database lists the musicians that played in each
band without distinguishing which musicians played at different times,
so it is not possible to study the time evolution of the collaboration
network.  The bands contain $1275$ different names of musicians.
However, it is important to stress that this number does not
necessarily represent the number of individuals, similar to the
comment noted in \cite{Newman2}. The same musician might appear with
different names in the database. For example, the musician Henry Allen
appears as Henry Allen, Red Allen or Henry Red Allen.  Also in some
bands the names of some musicians are not known. As a consequence they
are cited in the database with the same name: {\it unknown}.  In
Fig. \ref{fig1} we show the distribution of musicians that have played
in a band. It presents a skewed shape, with a peak around $5-10$
musicians and a few large bands that include up to $171$ musicians.

Let us begin with the collaboration network of individuals. Since some
 musicians have played in more than one band we obtain a modular
 network, formed by fully connected modules (bands) which are also
 connected between each other. Using this definition we obtain a
 network with a mean distance of $2.79$ and, as we show later, with
 very high clustering. The bands network is also a small world, with a
 mean distance of $2.26$, slightly smaller than the distance between
 musicians. This is an expected result, since in the musicians network
 an individual may need to contact a musician in his band in order to
 reach a musician in another band.

\begin{figure}[!h]
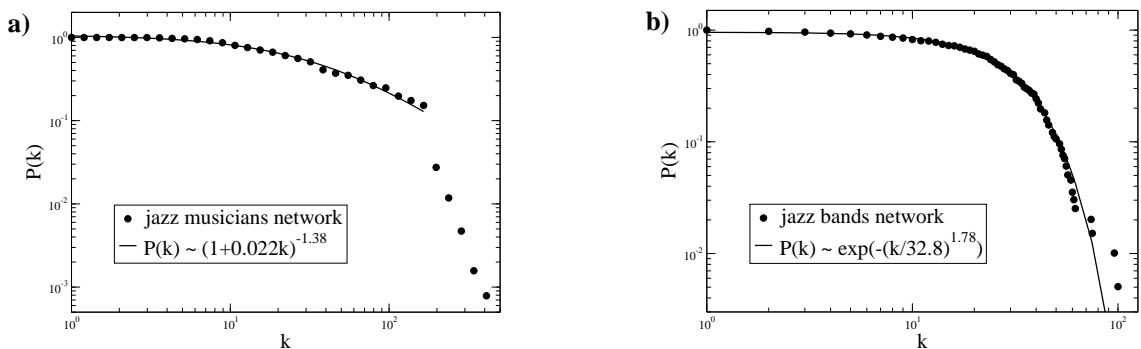

  \vspace{1cm}
  \centering{
    \includegraphics[width=0.4\columnwidth]{figure2a}
    \hspace{0.1\columnwidth}
    \includegraphics[width=0.4\columnwidth]{figure2b}
  }
  \caption{\label{fig2} Cumulative degree distribution $P(k)$ for the
    jazz musicians network (a) and the jazz bands network (b).}
\end{figure}

We measure the typical quantities used to characterize complex
networks. First we study the degree distribution and its
correlations. Then we look at the clustering coefficient and how it
depends on the degree $k$.  In Fig. \ref{fig2} we show the cumulative
degree distribution $P(k)$, i.e. the probability for a vertex to have
a degree larger or equal to $k$.  In the musicians network, shown in
Fig. \ref{fig2}(a), a slow decay with a power law tail $P(k) \sim
(1+a_0 k)^{-\alpha_m}$ with $a_0=0.022(1)$ and $\alpha_m=1.38(1)$ is
observed for values of $k$ up to $k=170$. This value coincides with
the degree of the musicians which have only played in Paul Whiteman's
Orchestra, the band that includes the largest number of musicians. For
$k>170$ a faster decay is observed. In table I we present
the names of the ten most connected musicians with their corresponding
$k$ values.

\begin{table}[!ht]
\title{The ten most connected musicians.}
{\begin{tabular}{@{}ccc@{}} \toprule
Position& Name& k\\ 
\hline
1          &      Eddie Lang            &   415\\
2          &      Jack Teagarden        &   387\\
3          &      Frank Signorelli      &   332\\
4          &      Frankie Trumbauer     &   307\\
5          &      Joe Venuti            &   295\\
6          &      Bob Mayhew            &   287\\
7          &      Bill Rank             &   267\\
8          &      Fud Livingston        &   265\\
9          &      Miff Mole             &   262\\
10         &      Louis Armstrong       &   262\\
\end{tabular}}
\label{table}
\caption{The ten most connected musicians with the corresponding degree}
\end{table}

Fig. \ref{fig2}(a) reveals that there are few musicians which have a
 very large number of connections.  Does this mean that these
 musicians play the role of hubs, and connect a large number of bands
 or is it simply that they have just played in the few bands which
 include the largest number of musicians?  One way to tackle this
 question is to eliminate all the links between musicians in the same
 band and consider only the links between bands. This is precisely
 what we do when we build the bands network.  In Fig. \ref{fig2}(b) we
 present the behavior of $P(k)$ for the bands network. A clear
 stretched exponential behavior $P(k) \sim exp(-(k/k^{*})^{\alpha_b})$
 with $k^{*}=32.8(1)$ and $\alpha_b=1.78(1)$ is observed. This result
 suggests that the bands are interconnected with a characteristic
 number of links. However we should keep in mind the fact that in the
 bands network construction two bands are connected by just one link,
 although they may share more than one musician. Clearly, just by
 analyzing the degree distributions we cannot answer the question. In
 order to advance one step forward we will study the correlations
 present in the networks.

\begin{figure}[!h]
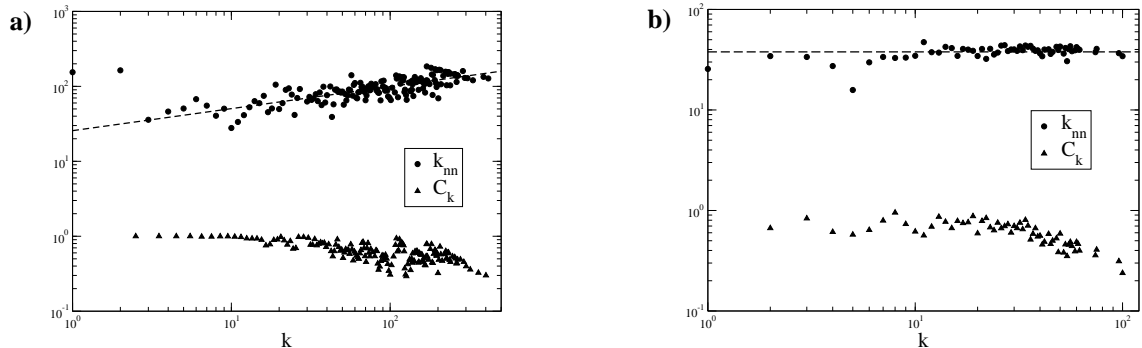

  \includegraphics[width=0.4\columnwidth]{figure3a.eps}
  \hspace{0.1\columnwidth}
  \includegraphics[width=0.4\columnwidth]{figure3b.eps}
  \caption{\label{fig3} Average degree $k_{nn}$ of nearest neighbors of
    vertices with degree $k$ (full circles) and clustering coefficient
    $C_k$ (full triangles) vs. $k$ for the musicians network (a), and the
    bands network (b).  The dashed lines are guides to the eye.}
\end{figure}

It has been observed that social networks show ''assortative mixing''
on their degrees, that is, there is a preference for high degree
vertices to connect to other high degree vertices \cite{Newman}. In
order to characterize the degree correlations present in the jazz
collaboration network we study the behavior of the average nearest
neighbors degree of the vertices of degree $k$ \cite{Pastor}, defined
as
\begin{equation}
k_{nn}(k) \equiv \sum_{k'} k'P(k'|k)
\end{equation}
where $P(k'|k)$ is the conditional probability that a vertex of degree
$k$ is connected to a vertex of degree $k'$. Fig. \ref{fig3}(a) shows
the behavior of $k_{nn}$ as a function of $k$ for the musicians
network. A clear increase in $k_{nn}$ is observed as $k$ grows.  For
$k < 170$ this could simply reflect the trivial fact that musicians
are already grouped in bands of different sizes. We known that the
musicians with $k>170$ have more links than the ones they get simply
by playing in one band. However, when we consider only these musicians
the behavior of $k_{nn}$ is very noisy, and no tendency can be
extracted.  Fig. \ref{fig3}(b) shows the behavior of $k_{nn}$ for the
bands network. In this case no correlation is observed, and $k_{nn}$
fluctuates around a constant value independent of $k$.  Again we
observe that both networks present different results which do not help
us to understand the role played by the most connected musicians.

We have also measured the clustering coefficient $C_k$ around nodes of
 degree $k$, which gives the fraction of neighbors of nodes with
 degree $k$ that are linked. Fig. \ref{fig3}(a) shows that the
 musicians network is highly clustered.  $C_k \simeq 1$ up to $k
 \approx 30$, and then presents a slow decay with oscillations. The
 peaks of these oscillations coincide with the $k$ values of the large
 bands. This result is expected by construction, since the musicians
 network is formed by fully connected modules which have $C=1$ by
 definition.  The bands network also presents a large clustering with
 $C_k \simeq 1$ up to $k \approx 30$ followed by a slow decay.

In this section we have analyzed the typical quantities which are used
to characterize the statistical properties of networks. The analysis
and comparison of the results obtained leaves a number of open
questions. On one hand the degree distribution of the musicians
network shows a skewed shape, with a power law tail and a small number
of musicians with a very large number of connections. On the other
hand the bands network presents a stretched exponential distribution
revealing that they are interconnected with a characteristic number of
links.  The correlations also show different properties, on one hand
the musicians network shows a positive correlation which seems to
indicate assortative mixing in the degrees, while the bands network
does not present any correlations at all.

How can we be confident that we are truly studying the collaboration
 network of jazz musicians? Are the different ways of constructing the
 networks creating misleading artificial artifacts or do they give us
 a real insight into the collaboration structure? In the next section
 we use the community structure analysis to answer these questions. As
 we show the different constructions do capture essential ingredients
 of the collaborations between the musicians. Also the community
 structure analysis allows for a quantitative characterization of the
 network other than the ones presented in this section.

\section{\label{sec3}Community structure of the jazz network}

To analyze the community structure we use the method proposed by
Girvan and Newman \cite{Girvan}.  The algorithm is based on the
concept of edge betweenness \cite{freeman}, which is defined as the
number of minimum paths connecting pairs of nodes that go through that
edge. The main idea is that edges which connect highly clustered
communities have a higher edge betweenness, and therefore cutting
these edges should separate communities. The algorithm proceeds as
follows:
\begin{enumerate}
\item The edge betweenness of every edge in the network is calculated
\item The edge with the highest betweenness is removed. If more than
  one edge has the highest value then one of them is chosen at random.
\item All the betweennesses are recalculated and step 2 is repeated
  until no edges remain.
\end{enumerate}

In order to describe the splitting process we generate a binary tree,
as described in \cite{Guimera}, where bifurcations depict communities
and leaves represent the actual vertices of the network. Note however,
that since the musicians network is constructed by connecting fully
connected modules, the tips of the branches do not necessarily
correspond to the most central musicians in the network. In fact, once
a band is disconnected from the large component all its edges are
completely equivalent.

\begin{figure}
  \centerline{\includegraphics[width=0.7\columnwidth]{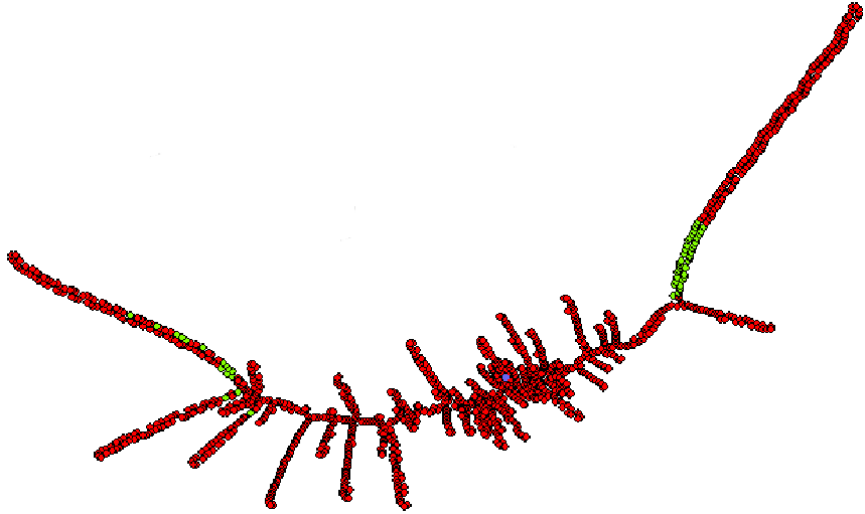}}
  \vspace*{8pt}
  \caption{\label{fig4} Community structure of the jazz musicians
    network. The root of the tree, in the middle of the figure, is
    indicated with the color blue.  The musicians with $k>170$ are
    indicated with green.}
\end{figure}

In Fig. \ref{fig4} we show the binary tree corresponding to the
musicians network. The root of the tree is indicated with a blue
circle. Going from the root of the tree to its tip, we note that first
a few small branches separate. A little further on, the tree splits
into two large branches or communities, which then split further into
smaller ones. This separation into two distinct communities can be
interpreted as the manifestation of racial segregation which was
present at that time. Although a small number of collaborations
existed between races, most bands were exclusively comprised of one
race or the other.  As a consequence a division in two large
communities separating black and white musicians should be present. In
fact, an analysis of the names of the musicians shows that the
musicians on the left community are black while the musicians on the
right are white.

In the previous section we have shown that the musicians network
presents a positive correlation on the degrees.  However, it was not
clear if this result indicated the presence of assortative mixing on
the degree or simply reflected the way in which the network was
constructed.  Let us analyze this point using the community structure
results. In Fig. \ref{fig4} we indicate with color green the musicians
with $k>170$. A visual inspection of the figure reveals that these
musicians are divided mainly into two communities. An analysis of the
names of these $55$ musicians \cite{Gleiser} reveals that the
musicians in the three communities on the left of the figure are only
black musicians, while the musicians in the right community are all
white. This results shows that the network presents assortative mixing
on the degree, and also shows a clear racial segregation at least for
the most connected musicians.

\begin{figure}
  \includegraphics[width=0.7\columnwidth]{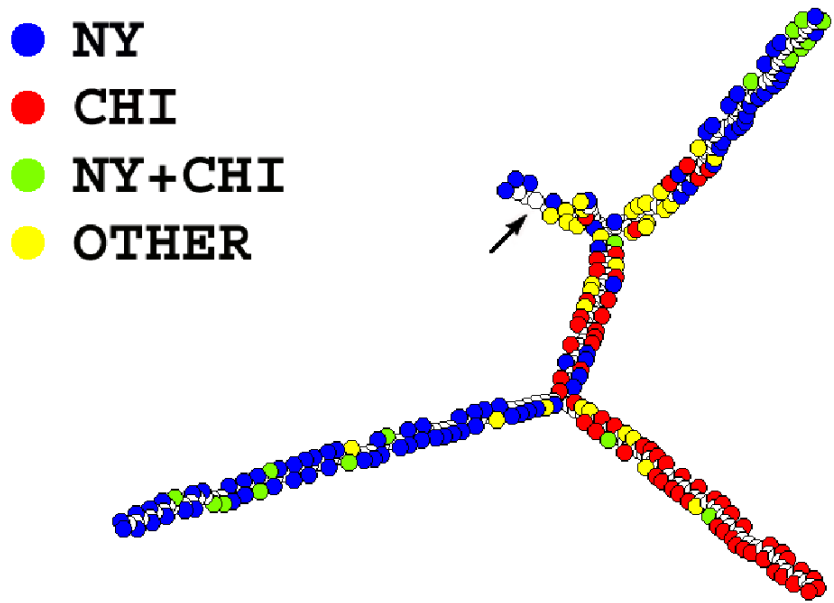}
  \vspace*{8pt}
  \caption{\label{fig5} Communities in the jazz bands network. The arrow
    indicates the root of the tree.  The different colors correspond to
    cities where a band has recorded: New York (blue), Chicago (red), both
    in New York and Chicago (green) and other cities (yellow).}
\end{figure}

Let us analyze now the community structure of the bands network.  In
Fig. \ref{fig5} we show the binary tree obtained using the Girvan
Newman algorithm. The bands network reveals a very simple community
structure. The tree is roughly divided into two large communities as
expected. However, the largest community also splits into two
communities. In order to show the origin of this division we have
analyzed the cities where the bands recorded. In Fig. \ref{fig5} we
indicate with color red the bands that have recorded in New York. The
bands that recorded in Chicago are indicated with blue. Clearly
Chicago and New York dominate as preferred recording locations. The
figure shows that the recording locations play a fundamental role in
the division of communities. Note that New York appears in two
different branches as a preferred recording location.  A closer look
at the names of the bands in the different communities reveals again a
racial segregation between black and white bands. The upper branch
correspond to bands with white musicians, while the lower branches
correspond to bands with black musicians.

\begin{figure}[!ht]
   \centerline{\includegraphics[width=0.6\columnwidth]{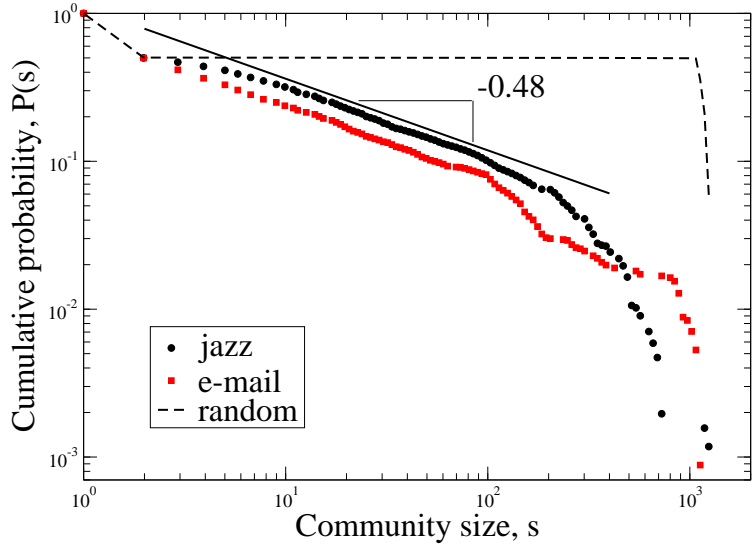}}
   \vspace*{8pt}
   \caption{\label{fig6} Community size distribution $P(s)$ as a
     function of community size $s$. The results for the jazz
     musicians network are plotted in full circles, while full
     triangles correspond to the e-mail network of University Rovira i
     Virgili. The dotted line corresponds to the results obtained in a
     random network with the same degree distribution as the musicians
     network.}
 \end{figure}

Let us resume the results we obtained in this section.
 Fig. \ref{fig4} clearly shows that the musicians network is
 assortative on the degree. It also reflects racial segregation
 between musicians.  We have also seen that the division into
 communities presents a strong correlation with the geographical
 locations where the bands recorded.  These results show that the
 musicians and bands network capture essential ingredients of the
 collaboration network of jazz musicians. With this in mind we
 characterize quantitatively the community structure.  In order to do
 so we calculate the cumulative community size distribution $P(s)$ as
 in \cite{Guimera}. This quantity gives the probability of a community
 to have a size larger or equal to $s$. Fig. \ref{fig6} shows that
 this distribution is heavily skewed, with a slow decay for community
 sizes up to $s \sim 200$, which roughly corresponds to the size of
 the largest band. This is followed by a faster decay and a cutoff
 corresponding to the size of the system at $s \sim 1000$.  A
 comparison of the shape of $P(s)$ with the results obtained in a
 recent study of the e-mail network of University Rovira i Virgili
 \cite{Guimera} shows a surprising similarity.  In fact, as shown in
 Fig. \ref{fig6}, both networks present a power law decay with the
 same exponent $-0.48$ up to $s \sim 200$. For small values of $s$ the
 jazz network deviates from this behavior, reflecting the fact that
 musicians are already grouped in bands of a certain size (see Fig
 \ref{fig1}), an effect which is not present in the e-mail network.

This result leaves a number of open questions. Are there fundamental
laws regarding the social interactions that lead to the formation of
community structures? Can these structures be characterized
quantitatively?  We hope that future work along these lines will help
us to answer these interesting questions.

\section*{Acknowledgments}

The authors acknowledge financial support from the Spanish Ministerio
de Educaci\'{o}n, Cultura y Deporte, DGES (Grant No.  BFM2000-0626)
and European Commission - Fet Open project COSIN IST-2001-33555. They
wish to thank Fede Bartumeus, Mari\'an Bogu\~n\'a, Albert D\'{\i}az
Guilera, Conrad P\'erez and Enric V\'azquez for useful comments and
suggestions, as well as Roger Guimer\`a and Alex Arenas for the
original code.  P. M. G. thanks Fundaci\'on Antorchas and L. D. thanks
Generalitat de Catalunya for financial support.

\end{document}